\documentclass[prl,twocolumn,preprintnumbers,amsmath,amssymb,floatfix]{revtex4-1}
\usepackage[dvipsnames]{xcolor}
\usepackage{soul}
\usepackage{amssymb}
\sethlcolor{green}

\usepackage{graphicx}% Include figure files
\usepackage{dcolumn}% Align table columns on decimal point
\usepackage{bm,textcomp}% bold math
\usepackage{booktabs,threeparttable}
\usepackage{subfigure}
\usepackage{epstopdf}
\usepackage{color}
\makeatletter % the following 3 lines change section titles to lowercase
\def\@hangfrom@section#1#2#3{\@hangfrom{#1#2#3}}
\makeatother
\makeatletter % the following 4 lines make [] style for reference
\def\@biblabel#1{[#1]}
\makeatother

\newcommand{\cm}[1]{} % comment
\DeclareGraphicsExtensions{.eps,.mps,.pdf,.jpg,.png}
\definecolor{bg}{rgb}{0.75,.75,.67}
\graphicspath{{./}}

\newcommand{\x}{{\times}} % {\times} removes space next to \times

\begin{document}

\title{Temperature rise in shear bands in a simulated metallic glass}
\author{Chunguang Tang$^{1*}$, Jiaojiao Yi$^{1,2}$, Wanqiang Xu$^{1}$, Michael Ferry$^{1}$}

\affiliation{$^1$ School of Materials Science and Engineering, The University of New South Wales, NSW 2052, Australia; $^2$ State Key Laboratory of Metal Matrix Composites, School of Materials Science and Engineering, Shanghai Jiao Tong University, Shanghai 200240, PR China}

\begin{abstract}
Temperature rise ($\Delta T$) associated with shear-banding of metallic glasses is of great importance for their performance. However, experimental measurement of $\Delta T$ is difficult due to temporal and spatial localization of shear bands and, as a result, our understanding of the mechanism of $\Delta T$ is limited. Here, based on molecular dynamics simulations we observe a spectrum of $\Delta T$, which depends on both sample size and strain rate, in the shear bands of CuZr metallic glass under tension. More importantly, we find that the maximum sliding velocity of the shear bands correlates linearly with the corresponding $\Delta T$, ranging from $\sim$25 K up to near the melting point for the samples studied. Taking heat diffusion into account, we expect $\Delta T$ to be lower than 25 K for the lower end of sliding velocity. At high temperature, shear band bifurcation and/or multiplication can occur as a negative feedback mechanism that prevents temperature rising well above the melting point.
\end{abstract}

\maketitle

\section{Introduction}
Shear-banding is a form of highly localized plastic deformation widely found in polymers, granular suspensions, and even crystalline metals \cite{li_behavior_1984,fall_yield_2009, wei_evolution_2002, xue_self-organization_2002}. It is also the major mode of plastic deformation in metallic glasses below their glass transition. During this process, the stored elastic energy of the strained material transforms into kinetic energy of the shear bands, which is expected to result in an increase in temperature within the bands. Once shear bands form in metallic glass, they can slide either via a runaway mode or serrated (stop-and-go) mode. The runaway sliding often leads to a catastrophic failure, and the serrated shear-banding can also transit into fracture \cite{das_temperature_2018, maass_long_2015}. Upon fracture, substantial temperature rise is observed, as evidenced by sparks during dynamic fracture \cite{gilbert_light_1999}, the formation of liquid droplets \cite{liu_test_1998}, and river patterns \cite{bruck_dynamic_1996} on fracture surfaces.

Intuitively, one would expect higher temperature rise in samples of larger size and at larger strain rates \cite{greer_shear_2013}. A modelling analysis, based on the assumption that the heat flux into the band scales with the sliding velocity, reports that the temperature rise $\Delta T$ is less than 15 K for samples with diameter up to 2 mm but is greater than the glass transition temperature for samples of at least 3 mm in diameter \cite{cheng_cold_2009}, and infra-red thermography of constrained compressions demonstrates that $\Delta T$ increases from $\sim0.5$ K to $\sim3$ K as the strain rate increases from 4.9$\x$10$^{-4}$/s to 2.8$\x$10$^{-2}$/s \cite{jiang_rate-dependent_2008}. Although $\Delta T$ measured by infra-red thermography typically ranges from 0.25 K to several tens of K  \cite{yang_-situ_2004, yang_dynamic_2005, song_seeing_2017, flores_local_1999},  $\Delta T$ $\sim$500 K during dynamic compression has been reported \cite{bruck_dynamic_1996}, albeit based on the use of a single detector. $\Delta T$ values around 1127 K upon dynamic fracture has also been reported based on light spectrum analysis \cite{gilbert_light_1999}. 

While the above studies confirm the rise in temperature associated with  shear-banding, direct and accurate measurements of $\Delta T$ in a shear band is nearly impossible because it occurs within the bulk of the material and is only accessible for post-mortem analysis \cite{song_seeing_2017}. The accuracy of surface temperature measurements (such as infra-red thermography) is limited by the temporal and spatial resolutions of these techniques, which are orders of magnitude lower than that required (in nm and ns to $\mu$s scales). As a result, $\Delta T$ is often indirectly estimated based on solutions to the heat conduction equation  or adiabatic approximation. As summarized in Table \ref{tab:temp}, such estimations can, even in the scenario of high resolutions of $\Delta T$ measurement \cite{lewandowski_temperature_2006}, lead to unrealistic temperature predictions up to 8600 K or more \cite{lewandowski_temperature_2006, yang_localized_2006, bengus_peculiarities_1993}.

As can be seen, the measurement of temperature within shear bands is intrinsically difficult, and our understanding of temperature rise is limited. On the other hand, atomistic simulations \cite{cao_structural_2009, bailey_atomistic_2006}, thanks to their nanoscale capacity comparable to the spatial and time scale of shear banding, are a promising complementary approach for tackling this difficult problem. Through simulating tensile loading of CuZr metallic glass we attempt to elucidate the mechanisms of heating associated with shear banding and contributing effects of sample size and strain rate. 

\begin{table*}
\tabcolsep=9pt
\caption{Summary of measured and calculated temperature rise in shear bands. IR, LS, and STZ denotes infrared imaging, light spectrum, and shear transformation zone, respectively.}
\begin{tabular}{l l l l l l l }
\hline \hline
material & deformation & $\Delta T_{\rm{measured}}$ & method & $\Delta T_{\rm{calculated}}$ & calculation method & reference\\  
\hline
Pd-based & tension & 0.04 & thermometer & $\gg$ 8000 & adiabatic approximation & \cite{bengus_peculiarities_1993}\\
Zr-based & tension & 0.25 & IR & 333 & STZ heating & \cite{yang_localized_2006}\\
~ & ~ & ~ & ~ & 10000 & adiabatic approximation & \cite{yang_localized_2006}\\
Zr-based & tension & 0.4 & IR & 1200 & adiabatic approximation & \cite{yang_-situ_2004}\\
Zr-based & tension & 22.5 & IR & 54.2 & heat conduction & \cite{flores_local_1999} \\
Zr-based & bending & $>$207 & tin coating & 3100-8300 & Carslaw-Jaeger's equation & \cite{lewandowski_temperature_2006}\\
Zr-based & compression & $\sim$500 & IR & - & - & \cite{bruck_dynamic_1996}\\
Zr-based & impact & $\sim$1127 & LS & - & - & \cite{gilbert_light_1999}\\
\hline\hline
\end{tabular}
\label{tab:temp}
\end{table*}

\section{Methods}
First, a bulk rectangular simulation cell containing 2592 Cu and 2592 Zr atoms, modelled by the embedded atom method (EAM) potential proposed for CuZr \cite{mendelev_development_2009}, was well equilibrated at 2000 K, about 700 K higher than the simulated melting point \cite{tang_anomalously_2013}, for 2 ns. The liquefied system was then quenched to 100 K at 10$^{10}$ K/s, with the final cell dimension being around $2\x 3.9\x 11.8~\rm{nm}^3$. A tension sample, denoted as $n_{x}\x  n_{y}\x  n_{z}$, was then built by repeating the final configuration by $n_{x}$, $n_{y}$, and $n_{z}$ times in space.  For reducing computational cost, we set $n_{x}$ as 1 for all samples in this study, which makes the size along $x$ much smaller than those along $y$ and $z$. Nevertheless, this does not alter our conclusions since the $x$ size (2 nm) is much larger than the cut-off (7.6 $ $\AA$ $) range of atomic interaction and periodic boundary conditions are applied along $x$. The size of simulation cell along $y$ was then doubled by inserting in a vacuum layer to create free surfaces. The samples with free surfaces were then relaxed at 650 K and 100 K for 80 ps and 20 ps, respectively, to reduce the structural periodicity caused by repeating the building blocks. Finally, a surface notch was created by removing some atoms in order to stimulate the formation of shear bands from the surface. The notched samples were then stretched along $z$ axis up to an engineering strain of $\varepsilon=0.2$ at a strain rate of 10$^8$/s unless otherwise specified. Structure visualization and analysis were performed using Code OVITO \cite{stukowski_visualization_2010}. 

All simulations were carried out under NPT (constant particle number, pressure, and temperature) ensemble using code LAMMPS \cite{plimpton_fast_1995}, with temperature and pressure being modulated with Nos\'{e}-Hoover thermostat and barostat, respectively, and the time step being set as 2 fs. During the bulk relaxation process, pressure was isotropically set to zero, and during the tension process, pressure was set to zero only in $x$ direction. In order to avoid the influence of thermostat on temperature during deformation, we set a damping coefficient longer than the deformation process, which effectively turns off the thermostat\cite{cao_structural_2009}. While alternatively one can apply the thermostat only to a small group of atoms far away from the shear band to ensure the heat transfer from the shear band to the matrix, we noticed that within our current approach the temperature of the atoms very far from the shear band is essentially stable, due to the large size of the system with respect to the shear band, even when the shear band reaches its peak temperature.

Unless otherwise specified, for all temperature calculations in this study, we refer to a group of atoms in the center of the shear band that initially form a parallelepiped-shaped zone with the edges along the loading axis being 3 nm. Although this means a relatively wider zone is used for $\Delta T$ computation in a smaller sample, we note this choice does not qualitatively alter our conclusions in this work. We calculated the temperature of zones of interest using
\begin{equation}
T=2E_k/(3k_b)
\label{eq:2}
\end{equation}
where $E_k$ is the average kinetic energy of the atoms in the zones and $k_b$ is the Boltzmann constant. In this formula, the portion of kinetic energy from translational movement of atoms along the shear-banding direction should be factored out since this inertial collective movement does not contribute to temperature rise. We noted that at the center of a shear band the collective translational velocity is nearly zero due to the opposite movements of either side of the band, although the sliding velocity, i.e., the relative velocities of neighboring atom layers, can be high. As shown in Table \ref{tab:tk}, the translational effect is small and so ignored in this work. 

\begin{table} 
\caption{Contributions to temperature of atomic velocity components along the shear band direction ($T_1$), normal to the shear band but parallel to sample plane ($T_2$), and normal to sample plane ($T_3$) . Each contribution is obtained from the corresponding kinetic energy component averaged over atoms and divided by $1.5k_b$. Three parallelepiped-shaped zones with various distances, $d$, to the shear band center are examined for two samples at $\varepsilon$=0.08. It can be seen that the translational effect, or the difference between $T_1$ and the average of $T_2$ and $T_3$, is small within a shear band.} 
\tabcolsep=5.5pt
\begin{tabular} {cccccc}
  \hline
  \hline
Sample  &  $d$ ($ $\AA$ $) & $T_1$ ($K$) & $T_2$ ($K$) & $T_3$ ($K$) & $T_1$/$T_{total}$ \\
\hline
$1\x 5\x 5$ & 0 & 66.8 & 67.0 & 66.8  & 33.3\%\\
 ~ & 71 & 39.5 & 38.8 & 40.3  &33.3\% \\
 ~ & 141 & 36.0 & 37.1 & 36.3  & 32.9\% \\
$1\x 30\x 60$  & 0 & 306.9 & 291.8 & 296.6  & 34.3\% \\
 ~ & 141 & 143.2 & 78.9 & 78.0   & 47.7\% \\
 ~ & 283 & 93.5 & 39.8 & 35.3   & 55.5\% \\ 
\hline
\hline
\end{tabular}
\label{tab:tk}
\end{table}

\section{Results}

\begin{figure}[ht]
\includegraphics[width=3.3in]{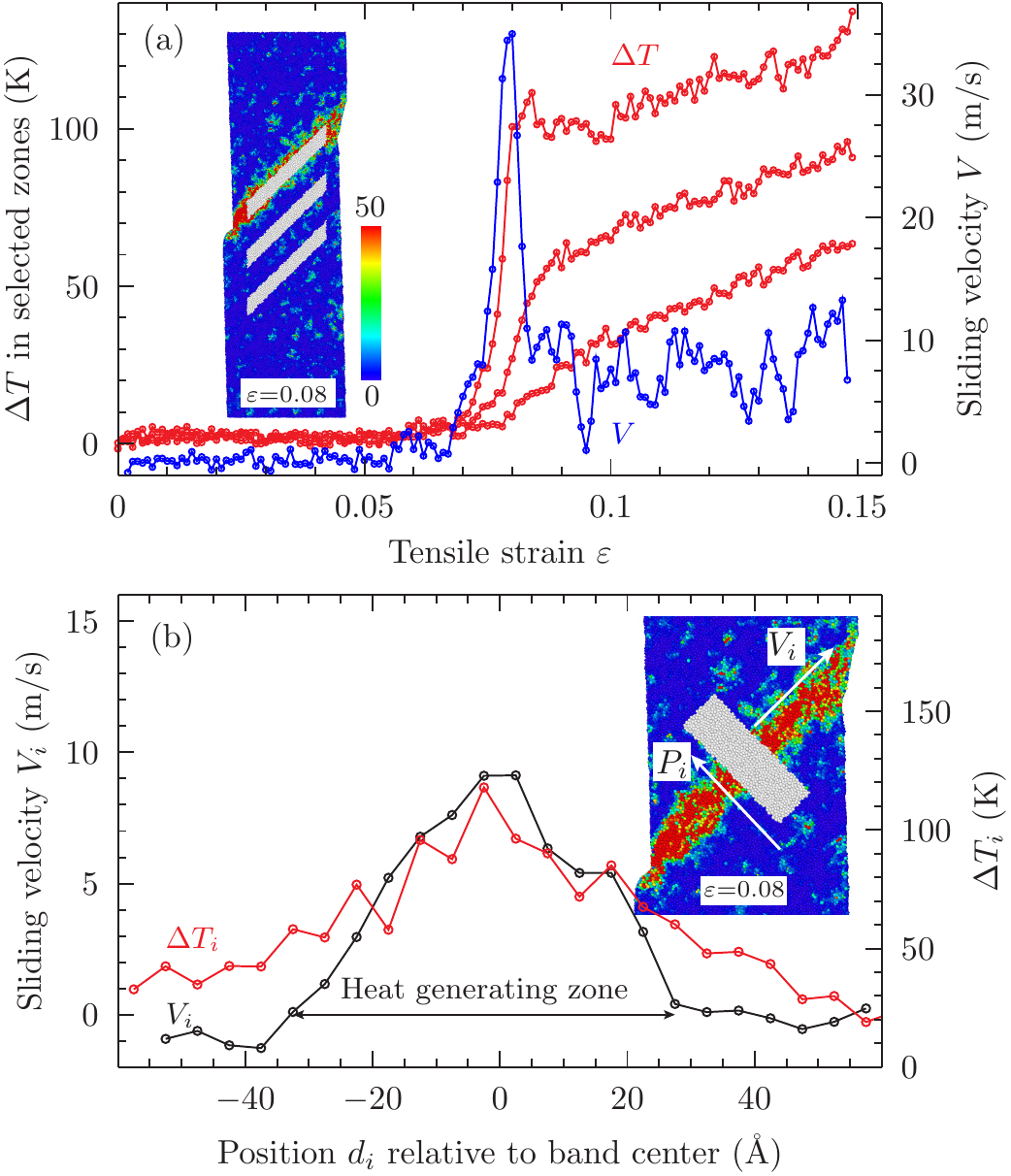}
\caption{Shear-banding in sample $1\x 5\x 5$. (a) Temperature rise $\Delta T$ in selected zones (highlighted in light grey in the inset) and sliding velocity of the shear band. $\Delta T$ is higher for a zone closer to the shear band. The color scale in the inset shows the non-affine squared displacement ($ $\AA$^2$) of atoms. For all figures, the sample images were viewed along $x$ axis. (b) Sliding velocity $V_i$ and temperature rise $\Delta T_{i}$ of layer $i$ within the selected zone (light grey in the inset) at $\varepsilon$=8\% as a function of its position $d_i$ relative to the center of the shear band. The positive directions of $i$ and $V_i$ are indicated by arrows.}
\label{fig:diffuse}
\end{figure}

\begin{figure*}[htb]
\centering
\includegraphics[width=6.4in]{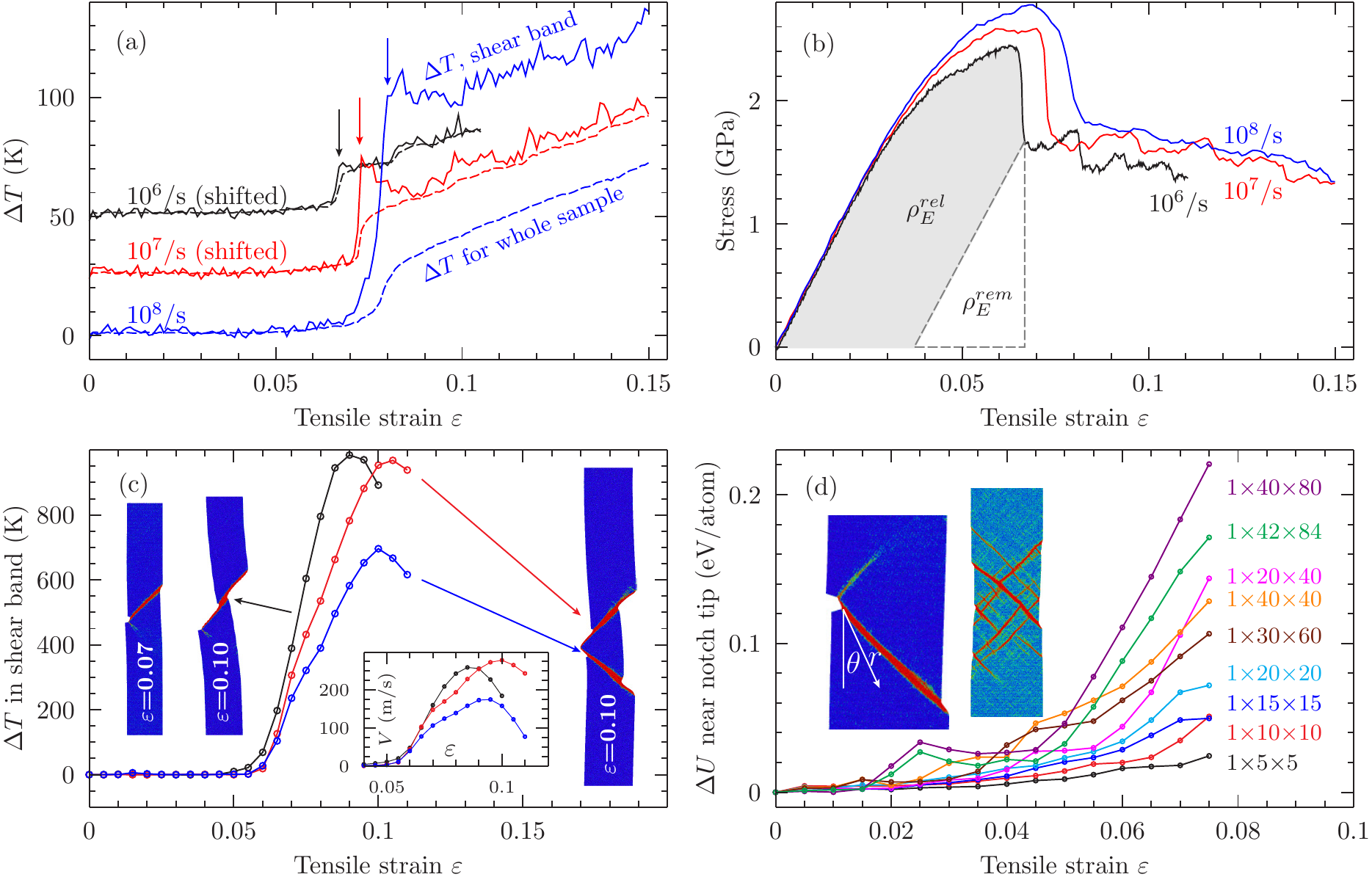}
\caption{(a) Temperature rise in the shear band for sample $1\x 5\x 5$ under various strain rates. $\Delta T$ is shifted by 25 K and 50 K for 10$^7$/s and 10$^6$/s, respectively. The dashed curves are for $\Delta T$ averaged over the whole sample. (b) Stress-strain curves corresponding to (a). The shaded area and the dashed triangle represent the density of elastic energy released ($\rho_E^{rel}$) and remaining ($\rho_E^{rem}$) when the shear band penetrates sample 10$^6$/s. $\rho_E^{rel}$ is  estimated to be 125, 104, and 81 MPa for 10$^8$/s, 10$^7$/s and 10$^6$/s, respectively. (c) Temperature rise in shear bands for samples $1\x 30\x 60$ (left inset) and $1\x 42\x 84$ (right inset). The inset plot shows the sliding velocities of the shear bands. (d) Internal energy change $\Delta U$ of a group of atoms next to the notch tip. The group is defined at $\varepsilon=0$ by a half-cylinder zone with $0\leq\theta\leq\pi$ and $r\leq$4 nm. The left inset, illustrating $\theta$ and $r$, is a local view of sample $1\x 40\x 40$ ($\varepsilon=0.07$) which produces an arrested immature shear band at $\theta\approx0.75\pi$. The right inset is a local view of multiple shear bands in sample $1\x 42\x 84$ without a surface notch.}
\label{fig:rate}
\end{figure*}

\begin{figure*}[thb] 
\centering
\includegraphics[width=6.9in]{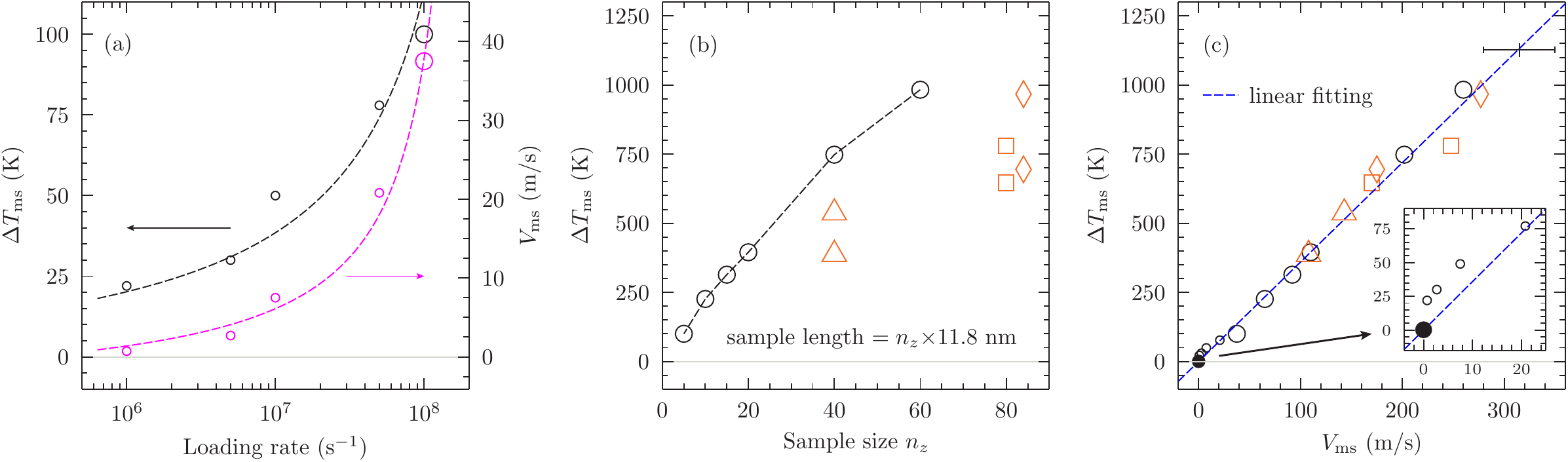}
\caption{(a) The maximum velocity of shear band sliding, $V_{\rm{ms}}$, and the corresponding temperature rise, $\Delta T_{\rm{ms}}$, as a function of strain rate for sample $1\x 5\x 5$. The loading rates lower than 10$^8$/s are represented by small circles. (b) Correlation between sample length and $\Delta T_{\rm{ms}}$ at fixed strain rate 10$^8$/s. The circles represent samples $1\x 5\x 5$, $1\x 10\x 10$, $1\x 15\x 15$, $1\x 20\x 20$, $1\x 40\x 40$, and $1\x 30\x 60$. The bifurcated bands are represented by $\triangle$ for sample $1\x 20\x 40$, $\square$ for $1\x 40\x 80$, and $\Diamond$ for $1\x 42\x 84$, respectively. (c) Correlation between $V_{\rm{ms}}$ and $\Delta T_{\rm{ms}}$. The solid circle and the point with an errorbar represent experimental data from combined sources as detailed in text. The dashed lines in (a) and (b) are for eye guide.}
\label{fig:size}
\end{figure*}

We start by examining the shear banding process associated with temperature rise using a $1\x 5\x 5$ sample as an example. For this sample, a mature shear band with thickness of $\sim$6 nm is formed at an engineering strain of $\varepsilon\approx0.08$, as shown in the inset of Fig. \ref{fig:diffuse}(a). 

As shown in Fig. \ref{fig:diffuse}(a), at $\varepsilon\approx0.07$ the temperature within the band zone starts to increase rapidly and reaches a peak $\Delta T\approx100$ K at $\varepsilon\approx0.08$. During this process the heat diffuses into the matrix and results in a temperature rise in the neighboring zones within the matrix. Fig. \ref{fig:diffuse}(a) shows that a peak of sliding velocity, or the relative velocity of either side of the shear band, occurs at $\varepsilon\approx0.08$, well corresponding to the sharp rise in temperature. After the peak, the sample does not fail catastrophically but the band continues to shear at around 5 to 10 m/s, corresponding to further temperature rise at larger strains. All these imply a close correlation between temperature rise and the friction of adjacent atoms within the shear band. To quantitatively elucidate this point, we chose the point of $\varepsilon=0.08$ and computed the sliding velocity of atomic layers (i.e., relative velocity of either side of a layer) as a function of their position relative to the center of the shear band. As shown in Fig. \ref{fig:diffuse}(b), the sliding velocity at the center of the band is the highest and decreases to zero at distances greater than ${\sim} 30~ $\AA$ $. The rise in temperature also exhibits a similar distribution, with $\Delta T>0$ at distances greater than ${\sim} 30~ $\AA$ $ due to heat diffusion. 

Since $\Delta T$ correlates with the sliding velocity  of shear band, strain rate is expected to directly affect $\Delta T$. As can be seen from Fig. \ref{fig:rate}(a), when the strain rate decreases from 10$^8$/s to 10$^7$/s and 10$^6$/s, the peak $\Delta T$ upon the point when the shear band penetrates the sample, as indicated by an arrow in the figure, reduces from ${\sim}100$ K to ${\sim}50$ K and ${\sim}25$ K. The trend observed here is qualitatively similar to that obtained using infrared thermography for quasistatic compression \cite{jiang_rate-dependent_2008} although the strain rate ranges in these two studies are orders of magnitude different.  During experimental deformation, especially dynamic testing, strain rate may not be constant. By step varying strain rate shortly before the start of shear banding, we found that it is indeed the strain rate accompanying shear banding, rather than the rate during the elastic deformation stage, that controls $\Delta T$. 

From an energetic perspective, $\Delta T$ results from the elastic energy released during the process of shear band formation. As shown in Fig. \ref{fig:rate}(b), the density of released energy somewhat decreases as strain rate decreases, but $\Delta T$ in the shear band decreases more rapidly. This implies that, at the point when the shear band penetrates the sample, for a lower strain rate a greater portion of kinetic energy diffuses from the band into the matrix. Indeed, as can be seen from Fig. \ref{fig:rate}(a), for rate 10$^6$/s $\Delta T$ in shear band is very close to that of the whole sample, which is distinct from the cases of the other two rates. This implies that for strain rate 10$^6$/s the sample is too small compared with the heat diffusion distance, resulting in near zero temperature gradient in the sample and, hence, low efficiency for further heat dissipation. It is clear that, for the same sliding velocity corresponding to sample $1\x 5\x 5$ at 10$^6$/s, one can expect even lower $\Delta T$ in a larger sample because of more efficient heat dissipation.

Using a constant strain rate of 10$^8$/s, we computed $\Delta T$ as a function of sample size. For samples up to $1\x 10\x 10$, the measured $\Delta T$ values are less than $\sim$200 K, comparable to previous simulation results \cite{cao_structural_2009, bailey_atomistic_2006}. As the sample size increases, we found a consistently increasing $\Delta T$, reaching $\sim$1000 K for sample $1\x 30\x 60$ of which the size is $\sim$$2\x 118\x 705$ nm$^3$, as shown in Fig. \ref{fig:rate}(c). Such an extreme temperature rise indicates that the shear band can reach the melting point ($\approx$1340 K \cite{tang_anomalously_2013}) of this simulated metallic glass during deformation at room temperature. As mentioned above, for sample $1\x 5\x 5$ a peak of $\Delta T$ occurs when the band propagates through the sample. However, for sample $1\x 30\x 60$ $\Delta T$ increases smoothly around this point. This indicates a transition from stable shear-banding to unstable shear-banding when the sliding velocity is larger than some threshold value, as was also observed in a model analysis \cite{cheng_cold_2009}.

Further increase in sample size beyond $1\x 40\x 80$ does not result in higher $\Delta T$. Instead, the shear band was found to bifurcate into two bands reducing the temperature, as shown in Fig. \ref{fig:rate}(c) for sample $1\x 42\x 84$. As shown in the inset of Fig. \ref{fig:rate}(c), the peak sliding velocities of the shear bands in samples $1\x 30\x 60$ and $1\x 42\x 84$ are about 260-280 m/s. It has been proposed that the sliding speed limit at the point of yield onset is about 1/10 \cite{miracle_shear_2011, greer_shear_2013} of the transverse sound wave speed, which ranges from $\sim$2 to $\sim$2.5 km/s for a series of BMGs \cite{wang_bulk_2004}. We note that the peak sliding velocities occur upon substantial plastic deformation and nearby the onset of yield the sliding velocities are far below the proposed limit. This indicates that $\Delta T$ can be limited by shear band bifurcation before the sliding velocity limit is reached, although we stress that bifurcation could occur at low-temperature shear bands due to, for example, local structural inhomogeneity. We note that a previous model \cite{cheng_cold_2009} also predicted higher temperature rise for larger samples, but in that model the temperature diverges and increases without limit \cite{greer_shear_2013} even upon small plastic strains for samples above some critical size. Consistent with our findings, experimentally, crack bifurcation around surface notch \cite{lowhaphandu_fracture_1998, flores_enhanced_1999} and multiple shear-banding \cite{conner_shear_2003} have been reported to increase fracture toughness and plasticity.

For the samples studied we found that shear band bifurcation is less likely for samples with $n_z$$\leq$20, likely for 40$\leq$$n_z$$\leq$60, and almost inevitable for $n_z$$\geq$80. Correspondingly, the change in internal energy, $\Delta U$, near the surface notch increases with increasing sample size, as shown in Fig. \ref{fig:rate}(d). This implies the bifurcation may result from the instability of shear band initiation driven by the large energy flux. This conclusion is supported by the occurrence of multiple shear bands in larger samples ($1\x 42\x 84$ and $1\x 30\x 90$) without a surface notch.

The effects of strain rate and sample size on temperature rise in shear bands can be illustrated using $\Delta T$ at some representative strain. Here we choose the point of the maximum sliding velocity ($V_{\rm{ms}}$), where the corresponding temperature rise, denoted as $\Delta T_{\rm{ms}}$, is near a peak or the maximum. For stable shear-banding (i.e., slow sliding), we did not choose the higher $\Delta T$ values during the following slower sliding process because these higher $\Delta T$ values result from the accumulated heating during the continuous sliding process while for serrated shear-banding the generated heat is nearly exhausted during the `stop' stage \cite{thurnheer_time-resolved_2016}. As can be seen in Fig. \ref{fig:size}(a), as strain rate decreases logarithmically both $V_{\rm{ms}}$ and $\Delta T_{\rm{ms}}$ appear to asymptotically approach zero. Fig. \ref{fig:size}(b) shows that, at fixed strain rate 10$^8$/s, $\Delta T_{\rm{ms}}$ increases approximately linearly with sample length for the single-band samples, but reduces considerably when shear band bifurcation occurs. By combining the data points for various strain rates and sample sizes, we obtain a linear correlation between  $\Delta T_{\rm{ms}}$ and $V_{\rm{ms}}$, as shown in Fig. \ref{fig:size}(c).

\section{Discussion}
Experimentally, measurements of shear-banding focus on the quasistatic cases that usually associate with serrated flow stresses \cite{dalla_torre_stick-slip_2010, song_capturing_2010, maass_propagation_2011, wright_high-speed_2013, slaughter_shear_2014} and generally the reported sliding velocities are on the order of 1 mm/s.   Infrared imaging \cite{yang_dynamic_2005, song_seeing_2017} reveals a temperature rise of $\sim$0.25 K corresponding to the drop phases of the stress serrations. These experimental data (0.25 K and 5.5 mm/s from references \cite{yang_dynamic_2005, song_capturing_2010}) are located on the lower-end extrapolation of our results (Fig. \ref{fig:size}(c)). Indeed, the recently reported $\Delta T$ up to 2 K is comparable to our lowest simulated $\Delta T$ ($\sim$20 K) when considering the sliding velocity difference.  Although flow stress serrations are beyond the capacity of MD simulations \cite{tang_atomistic_2018}, some insightful comments can be made here. As mentioned above, $\Delta T$ is limited by the elastic energy released during the shearing process. For serrated shear-banding, the stress drop is often below 0.04 GPa \cite{dalla_torre_stick-slip_2010, maass_shear-band_2015}, which means the released strain energy is only about 2\% of the fracture energy if a fracture stress of 2 GPa is assumed. If we assume $\Delta T$ around 1000 K for fracture \cite{gilbert_light_1999}, the possible maximum $\Delta T$ for serrated shearing would be 20 K. In view of the time scale difference between fracture and serrated shear-banding ($\mu$s versus ms \cite{slaughter_shear_2014}, $\Delta T$ for the latter should be far lower than 20 K.

On the other hand, the sliding velocity can be high during fracture. In a recent quasistatic compression experiment \cite{slaughter_shear_2014}, the fracture process of samples with dimensions $10\x 5\x 4.5$ mm$^3$ finished within $\sim$20 $\mu$s, which approximately translates into an average sliding velocity around 280 to 350 m/s ($\sqrt{2}\x 4.5/20$ to $\sqrt{2}\x 5/20$ mm/$\mu $s where $\sqrt{2}$ accounts for the shear band angle to the loading axis) although fracture might not be a complete shearing process. In this experiment, temperature rise was calculated to be 4060 K based on thermal diffusion equation, which appears to be overestimated since, for impact fracture, the temperature rise was measured to be about 1127 K based on light spectrum \cite{gilbert_light_1999}. Assuming $\Delta T$=1127 K and sliding velocity of 315$\pm$35 m/s upon fracture, we found this combination well fits the trend line in Fig. \ref{fig:size}(c). When comparing the fracture $\Delta T$ with our simulations, we are aware of the following two facts. First, we assumed an average sliding velocity (315 m/s) since the peak velocity was not available. Obviously, if the peak velocity was used, the corresponding data point should somewhat deviate from the fitting line. Second, fracture  is not an ideal shear-banding process and so can only be compared roughly with shear-banding. In fact, high speed imaging has shown that a fractured sample can break apart before the (measured) stress drops to zero \cite{wright_high-speed_2013}. With these said, the high temperature and fast velocity of fracture still seem to support the positive correlation between $\Delta T$ and sliding velocity.

Although so far no velocities are reported for the intermediate range in Fig. \ref{fig:size}(c), it is expected that during shock compression tests \cite{bruck_dynamic_1996, zhuang_shock_2002} and ballistic impacts the shear-banding is much faster than the quasistatic ones, which is worth for future experimental confirmation. Also, in the model analysis \cite{cheng_cold_2009} by Cheng $et~ al$., the sliding velocity for ``hot'' run-away shear-banding is far beyond the scale of mm/s. 

Finally, we note that the results obtained in this study are based on some specific relaxed structure (quenched at 10$^{10}$ K/s), specific environmental temperature (100 K), and a limited number of sample geometries. For lowering the computational cost, we used a thin slab geometry with only 2 nm in the $x$ direction. While this thickness is much larger than the cutoff distance of atomic interaction and indeed we obtained similar $\Delta T$ for a few thicker slabs, such as sample $2\x 5\x 5$, it is worth exploring the effect of sample geometry in the future. Also, as shear-banding is a thermally activated process \cite{maass_propagation_2011, derlet_thermally-activated_2018}, the effects of relaxation is worth further investigations. According to this study, ultrahigh temperature in the order of thousands of degrees in shear bands or fracture surfaces seems unlikely due to band or crack bifurcation. However, it remains as an open question whether this is possible in a scenario where the bifurcation is prohibited. In this case factors such as changes in viscosity may have an impact on the friction in shear bands and temperature rise.

\section{Summary}
In summary, we have studied temperature rise in shear bands during tensile loading of CuZr metallic glass through molecular dynamics simulations. It was found that the temperature rise correlates positively with strain rate and sample size. Especially, the maximum sliding velocity of the shear band, a function of both strain rate and sample size, correlates linearly with the corresponding temperature rise, ranging from $\sim$20 K to near the melting point. For quasistatic loading temperature rise lower than 20 K is expected from our study. Temperature well above the melting point was not observed because of shear band bifurcation or multiplication, which occurs in response to the instability of shear bands, hinders further temperature rise.

\section{Acknowledgements}
The authors acknowledge NCI National Facility for computational support of project codes eu7 and y88. CT would particularly like to thank the Australian Research Council for the DECRA Fellowship (grant no. DE150100738) for enabling this work to be carried out. JY thanks SJTU-UNSW Collaborative Research Fund for the financial support.
% \section{Competing Interests}
% The authors declare no competing interests.
% \section{Contributions}
% CT designed and performed the computations. All authors analyzed the data, discussed the results, and contributed to manuscript writing. CT wrote the manuscript.

\section{References}
% ********* main text ends here *************
%\input{text.bbl0}

\end{document}